\documentclass[12pt,thmsa]{article}
\usepackage{amsfonts}
\usepackage{graphicx}
\usepackage{epsfig}
\usepackage{graphics}

\makeatletter
\newcommand{\row}[1]%
{\mathord{\buildrel{\lower3pt%
\hbox{$\scriptscriptstyle\rightarrow$}}\over #1}}
\newcommand{\col}[1]{{#1^{\raisebox{2pt}[\height]%
{$\scriptstyle\downarrow$}}}}
\newcommand{\dyadic}[1]{\mathord{\dyadic@rrow{#1}}}
\newcommand{\dyadic@rrow}[1]{
\begin{picture}(12,12)(-1,0)
\put(-1,9){\makebox(0,0)[t]{$\scriptscriptstyle\downarrow$}}
\put(-1,9){\makebox(0,0)[l]{$\scriptscriptstyle\longrightarrow$}}
\put(5,0){\makebox(0,0)[b]{$#1$}}
\end{picture}
}

\newcommand{\bra}[1]{\bigl\langle #1 \bigr|}
\newcommand{\ket}[1]{\bigl| #1 \bigr\rangle}

\input{tcilatex}

\begin{document}
\begin{center}
{\large Single  and  double  changes of  Entanglement}\\[0pt]
\vspace{0.5cm} Nasser Metwally \\
 Math. Dept., College of Science, University of
Bahrain,\\ Kingdom of Bahrain\\
 Math. Dept., Faculty of science, Aswan,  University of Aswan,
\end{center}

\begin{abstract}
Entanglement behavior for different classes of two qubit systems
passing through a generalized amplitude damping channel is
discussed. The phenomena of sudden single,   double  changes and
the sudden death of entanglement are reported for correlated and
non- correlated noise.  It is shown that,  for less entangled
states, these phenomena appear for small values of channel
strength. The effect of the channel can be frozen for these
classes as one increases the channel strength. Maximum entangled
states are more fragile  than partial entangled states, where the
entanglement decays very fast. However, one can not freeze the
effect of  the noise channel  for systems initially prepared in
maximum entangled states. The decay rate of entanglement for
systems affected by non-correlated noise is much larger than that
affected by correlated noise.

\end{abstract}

pacs:03.65.-w,03.65.Ta,03.65.Yz,03.67.-a,42.50.-p


\topmargin=0.001cm \textheight=24cm \textwidth=17cm

\section{Introduction}
Entanglement represents an invaluable recourse for most of quantum
information tasks \cite{Nielsen}. It is possible to generate
maximum entangled states  but due to its interaction with
environment, the entanglement decreases and consequently the
efficiency of using it to perform  some quantum information and
computations decreases. Sometimes, the entanglement decays  during
sending the states from the lab to the users. However, there are
some cases where the travelling states are  forced to pass through
a noise channel \cite{Bandy}. Therefore it is important to
investigate and quantify the decay rate of entanglement due to
this undesired interactions.

There are some common noise channels that have been considered as
unavoidable noise during performing quantum information tasks. For
example, Yu and Eberly \cite{Eberly2004} showed that, the abrupt
and asymptotically gradual decay of entanglement  predicted in
amplitude and phase damping channels. However for some systems the
entanglement is completely vanishes in a finite time. This type of
entanglement decay  is called entanglement sudden-death, ESD
\cite{Eberly2006} and  has been reported extensively in many
studies \cite{metwally2010}.

It  has been shown that, there are some classes of noise states
whose teleportation fidelity can be enhanced if one of the two
qubits subject to dissipative interaction with environment via
amplitude damping channel \cite{Badziag2002}. The possibility of
recovering entanglement in the presence of amplitude damping
channel is discussed by Sun,  et.al \cite{Sun,Man}. Recently,
Montealegre et. al \cite{Mont2013}, have investigated the effect
of  different types of noise channels on the on-norm geometric
discord.

This motivated us to study the entanglement behavior of different
classes of maximum and partial entangled states. The current
manuscript is different from the previous manipulation
\cite{Mont2013}, where it is assumed that both qubit interact with
the noise channel. The noise could be correlated or
non-correlated. The behavior of entanglement shows that there are
different types of entanglement decays. The phenomena of sudden
change and sudden death of entanglement depends on the initial
state setting and the strength of the noise channel. However, the
effect of the noise channel can be frozen by a judicious choice of
the initial states and channel parameter.

The paper is organized as  follows: In Sec.II, we describe the
model of a two qubit-system. The initial degree of entanglement is
quantified for different classes via negativity. The generalized
amplitude damping and its effect on the two qubit system is
discussed in Sec.III, where we consider that the noise could be
correlated or non-correlated. In Sec. IV, we conclude  our
results.

\section{The suggested system}
Let us assume that there are two users, Alice and Bob, that
share a
 two-qubit  state. This state can be described
by $15$ parameters:  $3$  parameters, represent the Bloch vector
for each qubit and the $9$ remainder parameters represent the
correlation between the two qubits \cite{Nasser2000}. In these
parameters the two qubit state can be written as,
\begin{equation}\label{2QI}
\rho_{ab}=\frac{1}{4}(1+\row{s}\cdot\row\sigma^{(1)}+\row{t}\cdot\col\sigma^{(2)}+
\row{\sigma^{(1)}}\cdot\dyadic{C}\cdot\col\sigma^{(2)}),
\end{equation}
where $\row{s}=(s_x,s_y,s_z)$ and $\row{t}=(t_x,t_y,t_z)$ are the
Bloch vectors for the first and the second qubit respectively. The
vectors $\row\sigma^{(1)}=(\sigma_x^{(1)}, \sigma_y^{(1)},
\sigma_z^{(1)})$ and  $\row\sigma^{(2)}=(\sigma_x^{(2)},
\sigma_y^{(2)}, \sigma_z^{(2)})$ are the Pauli-vectors for each
qubit, respectively. The dyadic, $\dyadic{C}$  is $3\times 3$
matrix represents the correlation between the two qubits, where
$c_{ij}=tr\{\rho\sigma_k^{(i)}\sigma_l^{(j)}\}, i,j=1,2$ and
$k,l=x,y,z$.

The state (\ref{2QI}) is a general form of a two qubit systems,
where we can construct   several common classes in the context of
quantum information and computation. However, if we set
$\row{s}=\row{t}=0$, while the elements of the cross dyadic
$\dyadic{C}$ are $c_{ij}=0 $ for $i\neq j$ and $c_{ij}\neq 0$ for
$i=j$, one gets what is called $X-$state \cite{Eberly}.  The
Werner state  \cite{Werner} can be obtained from $X-$ state class
if we set $c_{11}=c_{22}=c_{33}=x$. The four Bell maximum
entangled states \cite{Hill} $\rho_{\phi^+}$, $\rho_{\phi^-}$,
$\rho_{\psi^+}$ and  $\rho_{\phi^-+}$ can be obtained for
different values of the parameters' $c_i$. For example  the
singlet maximum entangled state \cite{Nielsen, Hill}
$\rho_{\psi^-}$ can be obtained if we set
$c_{11}=c_{22}=c_{33}=-1$, where
$\rho_{\psi^-}=\ket{\psi^-}\bra{\psi^-}$ and
$\ket{\psi^-}=\frac{1}{\sqrt{2}}(\ket{01}-\ket{10})$.

\begin{figure}
\begin{center}
\includegraphics[width=20pc,height=15pc]{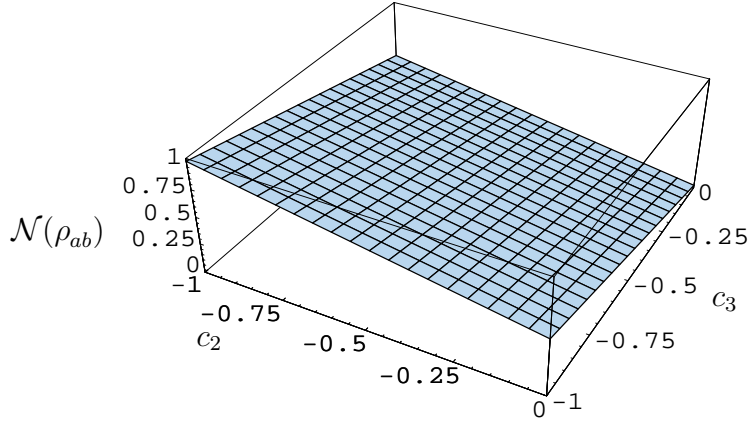}
\put(-210,40){$c_2$} \put(-15,55){$c_3$}
\put(-280,80){$\mathcal{N}(\rho_{ab})$}~
\end{center}
\caption{The negativity  $\mathcal{N}(\rho_{ab})$ as a measure of
entanglement for different classes of initial states with $c_1=-1
$ and  $c_2,c_3\in[-1,0]$  and  $p=\gamma=0$. }
\end{figure}

In our consideration, we  assume that the initial system is
prepared in $X-state$, Werner, or maximum entangled state. Since
the main aim of this contribution is investigating the behavior of
the amount of entanglement contained in these states after they
passing through a noisy channel, it is important to quantify the
initial amount of entanglement in these classes. For this purpose,
we use a measure of entanglement called negativity \cite{Kyc}.
This measure represents one of the most common measures of
entangled two qubit  states. This measure  based on positive
partial transpose criterion \cite{Peres}. Due to its operational
and calculations, negativity has been recently quantified
experimentally by Sliva et. al., \cite{Silva2013}. This measure
state that if $\lambda_i$ represent the eigenvalues of
$\rho_{ab}^{T_2}$, then the negativity is given by \cite{Kyc}
 \begin{equation}
 \mathcal{N}(\rho_{ab})=\sum_{k=1}^{4}{|\lambda_k|-1},
 \end{equation}
where $T_2$ represents a partial transposition for the second
qubit. As a function of the $c_i$ parameters, the negativity  can
be written as
\begin{equation}
\mathcal{N}(\rho_{ab})=-\frac{1}{2}+\frac{1}{2}tr\left\{\dyadic{C}^T\cdot\dyadic{C}\right\}.
\end{equation}
 For example the
degree of entanglement for Werner state is given by
$\mathcal{N}(\rho_{Werner})=-\frac{1}{2}+\frac{3|x|}{2}$ and
$\mathcal{N}(\rho_{Bell})=1$.
 In Fig.(1), we display the behavior of the negativity
$\mathcal{N}(\rho^{(f)})$ for different  classes of initial
states, where we set  $c_{xx}=-1$ and $c_{yy},c_{zz}\in[-1,0]$.
However, classes which are  described by $|c_{yy}| <1$ or
$|c_{zz}|<1$, represent partial entangled states, PES, where the
degree of entanglement is smaller than one, i.e., $0\leq
\mathcal{N}(\rho_{ab})<1$.

 \section{Noise channel}
The generalized amplitude damping channel, GAD represents one of
the most important channel in quantum information context. It
corresponds  to the interaction of a two-qubit system with a
squeezed thermal bath via a dissipative interaction
\cite{Strik2008}. Srikanth and Banerjee \cite{Strik2008} showed
that, the amplitude damping channel  preserves the non-classical
phenomenon from incoherence. This motivates  us to investigate the
behavior of the initial entangled  states passing through this
channel. The evolution of the  state (\ref{2QI}) under the effect
of the   generalized amplitude damped channel\cite{Mont2013} is
given by,

\begin{eqnarray}
\rho_{ab}^{(f)}&=&\sum_{i=0}^{3}
\Bigl\{\mathcal{U}_a^{(i)}\mathcal{U}_b^{(i)}\rho_{ab}\mathcal{U}_b^{\dagger
(i)}\mathcal{U}_a^{\dagger (i)}\Bigr\},
 \end{eqnarray}
 where
$\mathcal{U}_{a}$ and  $\mathcal{U}_{b}$  are the operators  of
the generalized amplitude damping channel for qubits $"a"$ and
$"b"$, respectively. In the computational basis, $"0"$ and $"1"$
these operators can be written as,
\begin{eqnarray}
\mathcal{U}^{(0)}_i&=&\sqrt{p}~(\ket{0}\bra{0}+\sqrt{1-\gamma}\ket{1}\bra{1})
,\quad \mathcal{U}^{(1)}_i=\sqrt{p}~\ket{0}\bra{1} \nonumber\\
\mathcal{U}^{(2)}_i&=&\sqrt{1-p}~(\sqrt{1-\gamma}\ket{0}\bra{0}+\ket{1}\bra{1}),\quad
\mathcal{U}^{(3)}_i=\sqrt{1-p}~\sqrt\gamma\ket{1}\bra{0},
\end{eqnarray}
where $i=a,b$,$\gamma=1-Exp[-\gamma_0 t]$ and  $\gamma_0$ is the
decay rate. It is clear that, if we set $p=1$ or  $0$, then the
generalized amplitude channel reduces to the usual amplitude
channel.
 The final state $\rho^{(f)}_{ab}$ depends on  the initial
state between the two users. In what follows we consider different
classes prepared initially with different degrees of entanglement.

 \subsection{ Correlated Noise}
 In this case, we assume that each qubit is effected by the same noise at the same time. For
 example,
 if the operator $\mathcal{U}_a^{(0)}$  is operators on the first qubit,
 then
 the second qubit will be  affected by the operator
 $\mathcal{U}_b^{(0)}$. Explicitly, the final state  evolves as,
\begin{eqnarray}
\rho_{ab}^{(f_c)}&=&\mathcal{U}_a^{(0)}\mathcal{U}_b^{(0)}\rho_{ab}\mathcal{U}_b^{\dagger
(0)}\mathcal{U}_a^{\dagger (0)}+
\mathcal{U}_a^{(1)}\mathcal{U}_b^{(1)}\rho_{ab}\mathcal{U}_b^{\dagger
(1)}\mathcal{U}_a^{\dagger (1)}
\nonumber\\
&&+\mathcal{U}_a^{(2)}\mathcal{U}_b^{(2)}\rho_{ab}\mathcal{U}_b^{\dagger
(2)}\mathcal{U}_a^{\dagger (2)}+
\mathcal{U}_a^{(3)}\mathcal{U}_b^{(3)}\rho_{ab}\mathcal{U}_b^{\dagger
(3)}\mathcal{U}_a^{\dagger (3)},
 \end{eqnarray}
where $\rho_{ab}^{(f_c)}$ is the final state in the presence of
correlated noise. In
 this case the final state (4) is described by

\begin{equation}
\rho_{ab}^{(f_c)}=\frac{1}{4}(1+ c_1\sigma^{(1)}_x\sigma^{(2)}_x+
c_2\sigma^{(1)}_y\sigma^{(2)}_y+ c_3\sigma^{(1)}_z\sigma^{(2)}_z),
\end{equation}
where,
\begin{eqnarray}
 c_1=\mathcal{B}_2+\mathcal{B}_3+\mathcal{B}_7+\mathcal{B}_8,
\quad
 c_2=- c_{1}, \quad
c_3=(\mathcal{B}_1+\mathcal{B}_6)-(\mathcal{B}_4+\mathcal{B}_5),
\end{eqnarray}
with
\begin{eqnarray}
\mathcal{B}_1&=&\frac{1+c_{zz}}{4}\left[p^2+(1-p)^2(1-\gamma)^2\right]
, \nonumber\\
\mathcal{B}_2&=&\frac{c_{xx}-c_{yy}}{4}\left[(1-\gamma)(1-p)^2+p^2(2-\gamma)\right],
\nonumber\\
\mathcal{B}_3&=&\frac{c_{xx}-c_{yy}}{4}(1-\gamma)\left[p^2+(1-p)^2\right]+
\gamma(1-p),
\nonumber\\
\mathcal{B}_4&=&\frac{1-c_{zz}}{4}(1-\gamma)\left[p^2+(1-p)^2\right]
, \quad
\nonumber\\
 \mathcal{B}_5&=&\mathcal{B}_4,\quad
\mathcal{B}_6=\frac{1+c_{zz}}{4}\left[p^2(1-\gamma)^2+(1-p)^2\right],
\nonumber\\
\mathcal{B}_7&=&\mathcal{B}_8=\frac{c_{zz}
+c_{yy}}{4}\left[p^2(1-\gamma)^2+(1-\gamma)(1-p)^2\right].
\end{eqnarray}
\begin{center}
\begin{figure}[t!]
\includegraphics[width=20pc,height=16pc]{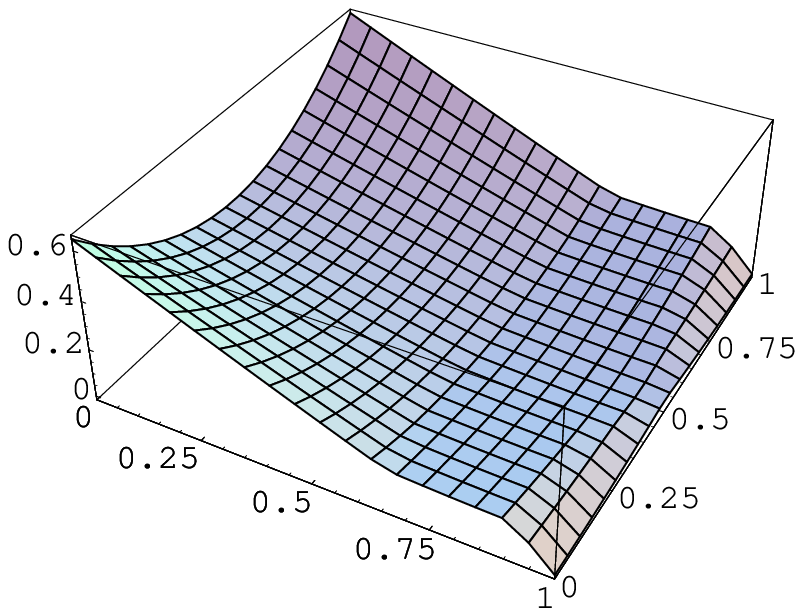}
\put(-160,20){$\gamma$} \put(-25,45){$p$} \put(-255,80){
$\mathcal{N}^{(f_c)}$}~\quad
\includegraphics[width=18pc,height=14pc]{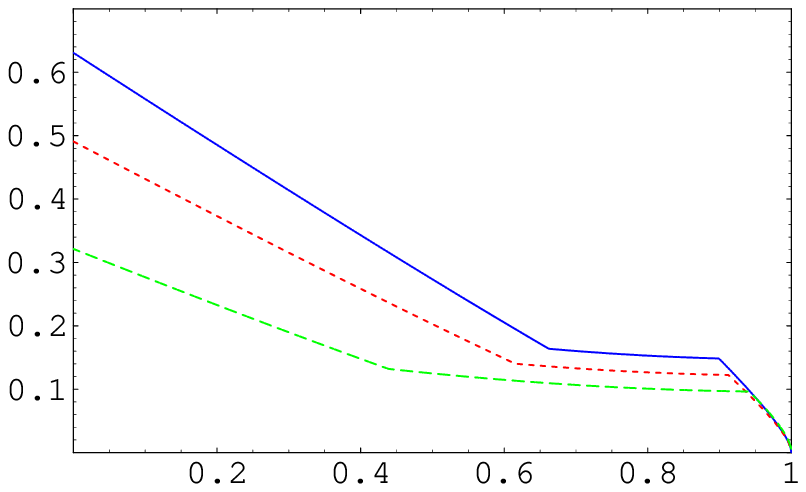}
 \put(-260,140){(a)}
\put(-25,150){(b)} \put(-240,90){ $\mathcal{N}^{(f_c)}$}
 \put(-100,-5){$\gamma$}~\quad
 \caption{The negativity
$\mathcal{N}^{(f_c)}$ against the channel parameters $p$ and
$\gamma$ for a state  initially described by $c_{xx}=-0.1,
c_{yy}=-0.2$ and $c_{zz}=-0.7$ and (b) the same as (a) but for
particular values of $p$. The solid, dot and dash curves for
$p=0.1,0.2$ and $0.3$, respectively.}
\end{figure}
\end{center}
\begin{figure}
\begin{center}
\includegraphics[width=18pc,height=14pc]{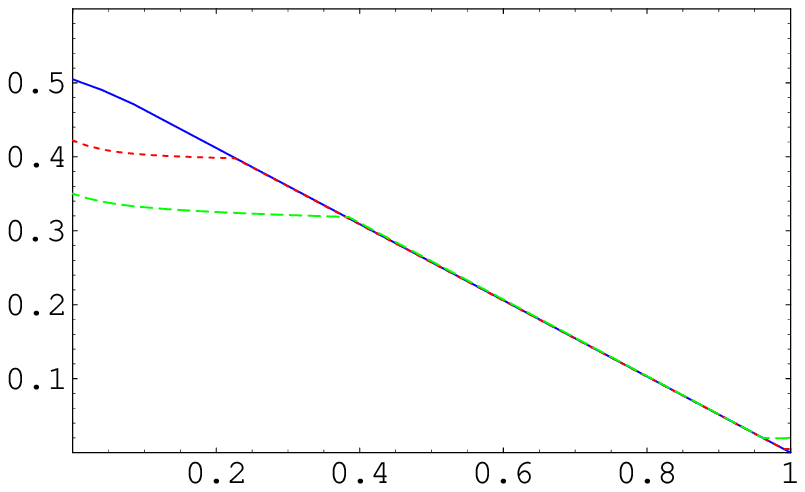}
\put(-120,-5){$\gamma$} \put(-250,98){
$\mathcal{N}^{(f_c)}$}~\quad
\includegraphics[width=18pc,height=14pc]{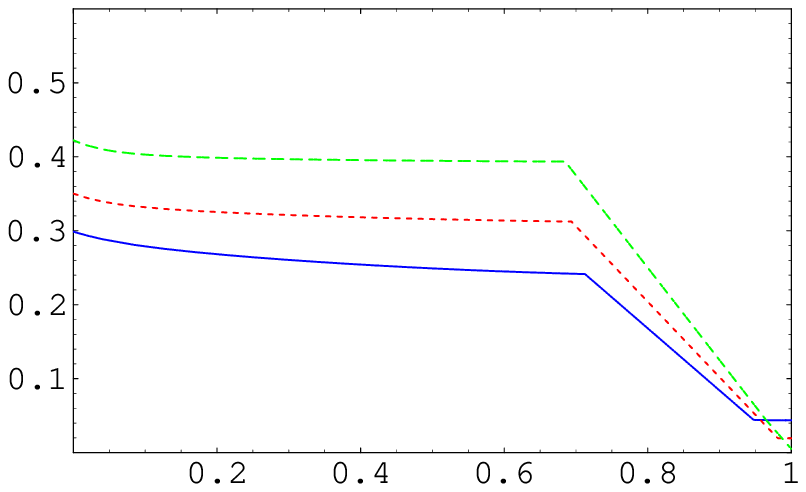} \put(-260,148){(a)}
\put(-25,150){(b)} \put(-100,-5){$\gamma$}
\put(-230,98){$\mathcal{N}^{(f_c)}$}
\end{center}
\caption{(a)The negativity  $\mathcal{N}^{(f_c)}$ for class
initially prepared in Werener state with
$c_{xx}=c_{yy}=c_{zz}=-0.03$. The solid, dot and dash curves for,
$p=0.01,0.1,0.2$, respectively and (b) the same as (a) but
$p=0.7,0.8,0.9$ for the  solid, dot and dash curves, respectively.
}
\end{figure}

The behavior of $\mathcal{N}(\rho^{(f_c)})$ for a  class of state
described by $c_{xx}=-0.1, c_{yy}=-0.2$  and $c_{zz}=-0.7$ is
described in Fig.(2a). From this figure three different phenomena
can be predicated. Firstly, there is a sudden decay of
entanglement as one increases the parameters $\gamma$ or $p$.
Secondly, there is a range of $\gamma$ and $p$, where the effect
of the noisy channel is completely frozen. Thirdly, a sudden death
of entanglement happens  at $\gamma=1$. In Fig. (2b), we
investigate the effect of a particular value of the channel
strength $p$ on the same system .  This figure describes clearly
the three previous phenomena. The upper bounds of entanglement
decrease as $p$ increases. As soon as $\gamma$ increases, the
entanglement decays sharply to reach its minimum bounds which
depend on the values of $p$. However, for larger values of
$\gamma$ the effect of the channel is frozen, namely, the
entanglement is immutable. The frozen interval increases  with
$p$:  the larger value of $p$ is, the larger frozen interval. The
last remark which appears in this behavior is the sudden
entanglement death   as  $\gamma$ further increases.

In Fig.(3), we consider a class of Werner state defined by
$c_{xx}=c_{yy}=c_{zz}=-0.03$. It is shown that, for small values
of $p$, the entanglement decays hastily to vanish completely at
$\gamma=1$. However for larger values of $p$, the entanglement is
almost constant  for $\gamma\in[0, 0.25]$, i.e., the effect of the
generalized amplitude channel is frozen. As $\gamma$ increases
further, the entanglement  suddenly decays to  vanish completely
at $\gamma=1$. The behavior of $\mathcal{N}$ shows that the frozen
interval of the channel increases as $p$ increases. The effect of
larger values of $p$ on the negativity of this class is depicted
in Fig.(3b), where we set $p=0.7,0.8$ and $0.9$. It is clear that,
the upper bounds of entanglement is smaller than those depicted in
Fig.(3a). Although the negativity decreases in general,  it
increases for larger values of $p$. The behavior of
$\mathcal{N}(\rho^{(f_c)})$ shows that the phenomena of the sudden
single, double changes  and channel frozen  appear for larger
values of $p$.

\begin{figure}[t!]
\begin{center}
\includegraphics[width=19pc,height=14pc]{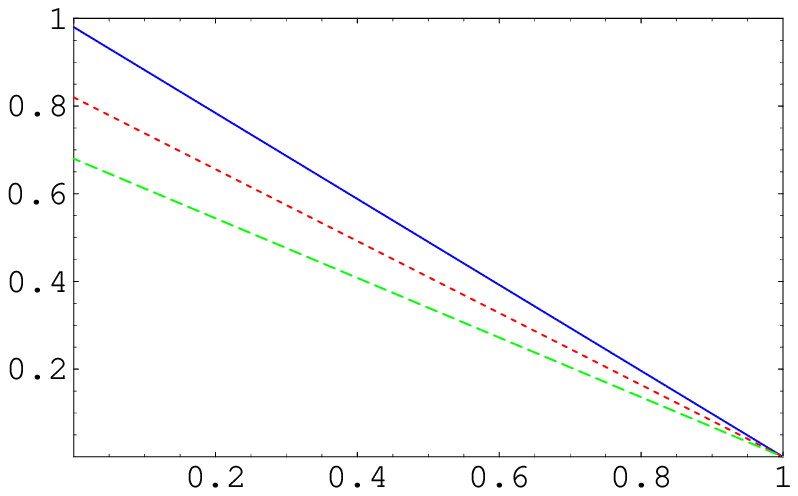}
\put(-115,-5){$\gamma$} \put(-245,88){$\mathcal{N}^{(f_c)}$}~\quad
\includegraphics[width=19pc,height=14pc]{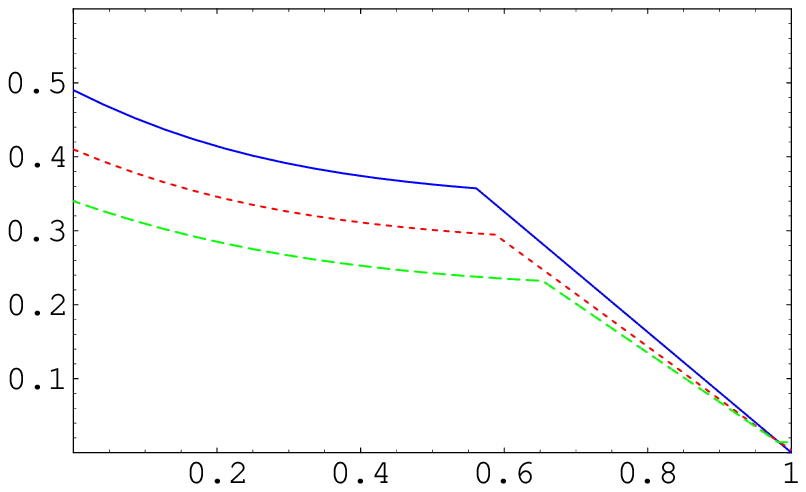} \put(-260,150){(a)}
\put(-25,150){(b)} \put(-100,-5){$\gamma$}
\put(-235,95){$\mathcal{N}^{(f_c)}$}
\end{center}
\caption{The same as Fig.(3) but the system is initially prepared
in (a) maximum entangled state i.e.,$c_{xx}=c_{yy}=c_{zz}=-1$(b)
in $X-$ state with  $c_{xx}=-0.1,c_{yy}=-0.2$ and $c_{zz}=-0.3$. }
\end{figure}

It is important to investigate the behavior of entanglement for
systems which are initially prepared in maximum entangled states.
 For this aim, we  consider a class defined by
$c_{xx}=c_{yy}=c_{zz}=-1$, where the entanglement $\mathcal{N}=1$.
Fig.(4a) shows that  the entanglement decays as $\gamma$ increases
and completely dies at $\gamma=1$. On the other hand, for larger
values of $p$, the upper bounds decrease.

Fig.(4b) displays the   entanglement  behavior of a class of $X-$
states, where we set $c_{xx}=-0.1, c_{yy}=-0.2$ and $c_{zz}=-0.3$.
It is clear that, the entanglement  slightly decays, then suddenly
decays to death completely at $\gamma=1$. The sudden changes
occurs faster for smaller values of $p$.

From the previous figures  one can conclude that, it is impossible
to freeze the noisy channel for systems prepared initially in
maximum entangled states. For less entangled state the phenomena
of sudden changes and sudden death appear for larger values of the
channel strength. However, the possibility of  the frozen noisy
channel increases for less entangled state and larger values of
the  channel's strength.

 \subsection{Non- correlated Noise}
In this subsection, we investigate the effect of the generalized
amplitude damping  channel for non correlated noise. In this case
the final state is given by,
\begin{equation}
\rho_{ab}^{(f_{nc})}=\sum_{i=0}^{i=3}\sum_{j=0}^{j=3}\Bigl
\{\mathcal{U}_a^{(i)}\mathcal{U}_b^{(j)}\rho_{ab}\mathcal{U}^{\dagger(j)}_b\mathcal{U}^{\dagger(i)}_a\Bigr\},
\end{equation}
where $\rho_{ab}^{(f_{nc})}$ is the final state for the initial
state (1) subject to non-correlated noise. This state can be
written explicitly  in the form (7) but  with different
coefficients $\tilde B_i, i=1,...8$   given by,
\begin{eqnarray}
\tilde{\mathcal{B}_1}&=&\frac{1+c_{zz}}{4}\Bigl(p+(1-p)(1-\gamma)\Bigr)^2
\nonumber\\
\tilde{
\mathcal{B}}_2&=&\frac{1-c_{zz}}{4}\Bigl[(1-\gamma)(p^2+(1-p)^2)+p(1-p)(1+(1-\gamma)^2)\Bigr]
,\quad \tilde B_3=\tilde B_2,
 \nonumber\\
\tilde{\mathcal{
B}_4}&=&\frac{1+c_{zz}}{4}\Bigl(p^2(1-\gamma)^2+(1-p)^2+p(1-p)(1-\gamma)\Bigr)+
\frac{c_{xx}-c_{yy}}{4}\gamma^2(1-p)^2,
\nonumber\\
\tilde
{\mathcal{B}_5}&=&\frac{c_{xx}-c_{yy}}{4}\sqrt{1-\gamma}(\gamma+p(1-\gamma)
 +\frac{c_{xx}+c_{yy}}{4}(1-\gamma)\Bigl[(2p-1)+p(1-p)(2+\gamma)\Bigr]
\nonumber\\
\tilde
{\mathcal{B}_6}&=&\frac{c_{xx}-c_{yy}}{4}\sqrt{1-\gamma}\Bigl(p^2+\gamma(1-p)^2+p(1-p)(1+\gamma)\Bigr)
\nonumber\\
&&\hspace{1cm} +\frac{c_{xx}+c_{yy}}{4}\Bigl[(1-\gamma)+\gamma
p(1-p)\Bigr],
\nonumber\\
\tilde
{\mathcal{B}_7}&=&\frac{c_{xx}-c_{yy}}{4}(1-\gamma)(1+p^2)+\frac{c_1+c_2}{2}p\sqrt{1-\gamma}
\nonumber\\
\tilde
{\mathcal{B}_8}&=&\frac{c_{xx}-c_{yy}}{4}(1-\gamma)(p^2+p-1)+\frac{c_1+c_2}{2}\gamma\sqrt{1-\gamma}(1-p)
\end{eqnarray}

\begin{figure}
\begin{center}
\includegraphics[width=20pc,height=16pc]{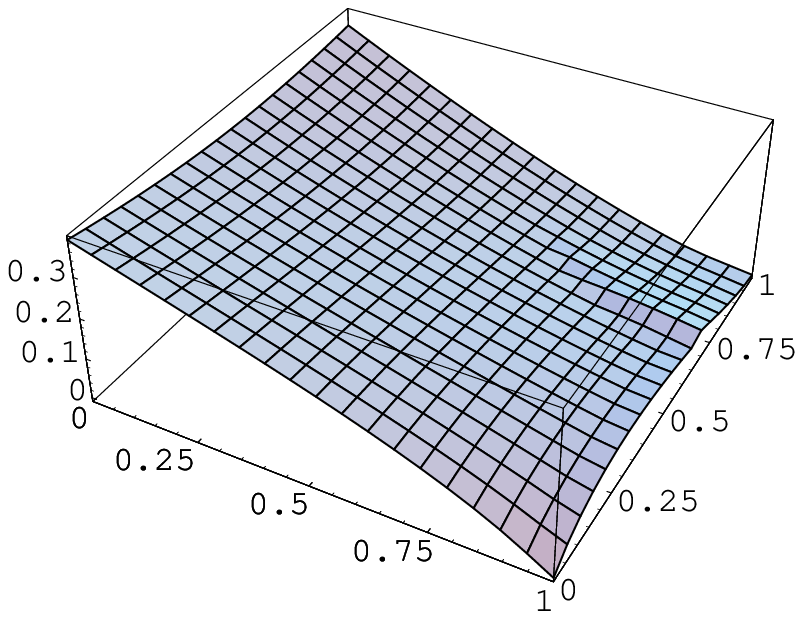}
\put(-160,20){$\gamma$} \put(-25,45){$p$} \put(-265,80){
$\mathcal{N}^{(f_{nc})}$}~\quad
\includegraphics[width=16pc,height=14pc]{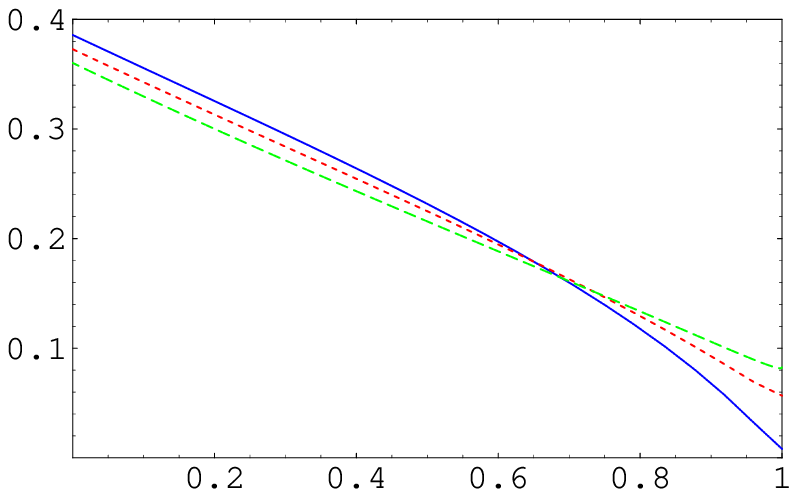}
\put(-250,140){(a)} \put(-25,140){(b)} \put(-90,-5){$\gamma$}
\put(-205,90){$\mathcal{N}^{(f_{nc})}$}
\end{center}
\caption{The same as Fig.(2), but it is assumed that the noise is
non-correlated.}
\end{figure}
In Fig.(5a), we investigate the effect of the non-correlated noise
on the same class that  is described in Fig.(2)  (correlated
noise). It is clear that, the upper bounds of entanglement are
smaller than those depicted for correlated noise (see Fig.(2a)).
As $\gamma$ increases the negativity $\mathcal{N}$ decreases to
vanish completely at $\gamma=1$.  At $\gamma=0$, the negativity
slightly decreases in the interval $p\in[0,0.75]$. However for
further values of $p$, there is a remarkable increase in the
negativity.

To describe the previous behavior of the negativity, we consider
different values of $p$ as shown in Fig.(5b). It is clear that,
initially ($\gamma=0$) the negativity decreases as $p$ increases.
For larger values of $\gamma\in[0,0.70]$, the negativity decreases
sharply. However for larger values of $\gamma$, the negativity
increases as $p$ increases.

\begin{figure}
\begin{center}
\includegraphics[width=16pc,height=14pc]{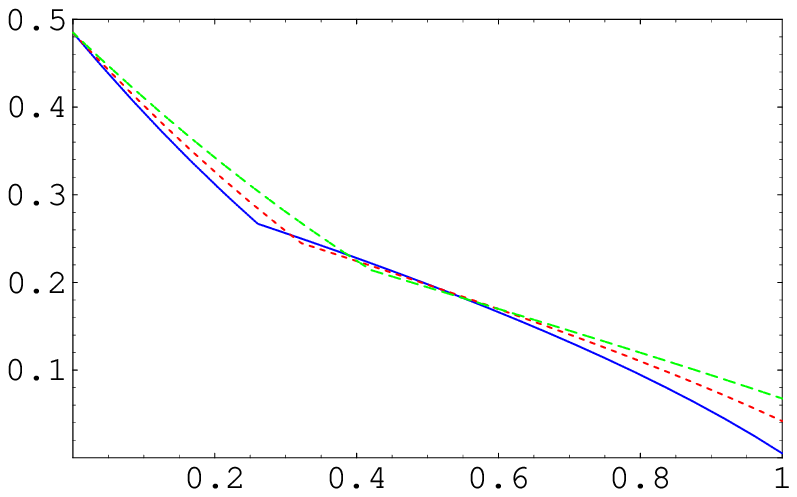}
\put(-90,-5){$\gamma$}
 \put(-210,85){$\mathcal{N}^{(f_{nc})}$}~\quad
\includegraphics[width=16pc,height=14pc]{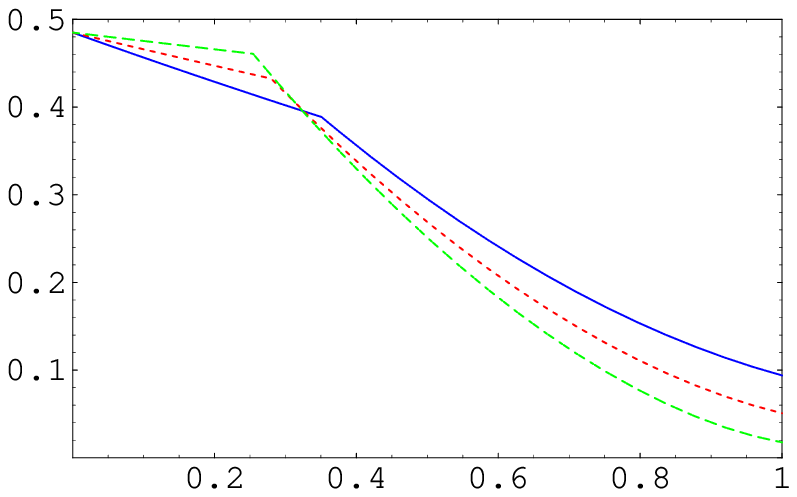}
\put(-250,145){(a)} \put(-25,145){(b)} \put(-90,-5){$\gamma$}
\put(-205,85){$\mathcal{N}^{(f_{nc})}$}
\end{center}
\caption{The Negativity for a system is initially prepared in a
Werner state defined by $c_{xx}=c_{yy}=c_{zz}=-0.3$ passes through
non-correlated noise. (a)~The solid, dot and dash curves for
$p=0.01,0.1,0.2$, respectively. (b)~The solid, dot and dash curves
for$p=0.7, 0.8.0.9$, respectively.}
\end{figure}

Fig.(6)  shows the behavior of  entanglement for a class is
initially prepared in Werner state  defined by
$c_{xx}=c_{yy}=c_{zz}=-0.03$ subjects to non-correlated noise. In
this case the  entanglement decays faster than that shown for
correlated noise (see Fig. 3). The phenomena of sudden decay and
sudden change appear clearly as $\gamma$ increases. However as $p$
increases the degree of entanglement increases and doesn't vanish
even at $\gamma=1$. The behavior of the negativity $\mathcal{N}$
for larger values of $p$ is described in Fig.(6b), where we set
$p=0.7,0.8$ and $0.9$. It is clear that, the rate of entanglement
decay is much smaller than that shown in Fig.(6a).

\section{conclusion}

We have studied the effect of the generalized amplitude damping
channel on  different classes of  two qubit systems. Particulary,
we considerd maximum entangled class, partial entangled class of
$X$-states and a class of Werner states. The degree of
entanglement is quantified for different classes of initial state
setting before passing  through the  generalized amplitude damping
channel.

 In this investigation, it is assumed that the noise could be correlated or non-correlated.
The general behavior of entanglement shows that, as the channel
parameters increase the entanglement decays. The  decay rate
depends on the initial state setting and  the noise type
(correlated or non-correlated). Our results shows that the decay
rate of entanglement is larger in the presence of the
non-correlated noise.

The phenomena of single,  double changes  and sudden death of
entanglement, as well as the frozen channel  appear according to
the initial state setting, the type of the noise and  the values
of the channel strength. For {\it correlated } noise,  if we start
from maximum entangled states, then  the entanglement decays very
fast and the decay's rate increases as the channel strength
increases. The phenomena of single changes and sudden death of
entanglement appear for systems prepared initially in Werner and
$X$-states with  small degree of entanglement. However, the sudden
double change of entanglement and the frozen channel phenomena
appear for systems prepared initially with small entanglement and
larger values of the channel strength.
 Although the entanglement decreases in the presence of
non-correlated noise, the upper bounds increase for smaller values
of the channel strength. However, for larger values of the channel
strength the entanglement decays smoothly to non-zero value

{\it In conclusion:}important phenomena might appear for two qubit
systems passing through correlated or non-correlated noise:
single, double changes, sudden death of entanglement and the
frozen noisy channel. One can increase the range of the  frozen
channel interval by controlling  the channel parameters. The
maximum entangled states are fragile while the partial entangled
states are more robust.

\end{document}